\documentclass[12pt]{article}
\usepackage[centertags]{amsmath}
\usepackage{amsfonts}
\usepackage{amssymb}
\usepackage{amsthm}
\usepackage{wasysym}
\usepackage{graphics}
\usepackage{setspace}
\newcommand{\beq}{\begin{equation}}
\newcommand{\eeq}[1]{\label{#1}\end{equation}}
\newcommand{\bea}{\begin{eqnarray}}
\newcommand{\eea}[1]{\label{#1}\end{eqnarray}}
\begin{document}
\newcommand{\Dirac}{/\!\!\!\!D}
\input FEYNMAN
\setlength{\topmargin}{-1cm} \setlength{\oddsidemargin}{0cm}
\setlength{\evensidemargin}{0cm}
\begin{titlepage}
\begin{center}
{\Large \bf Universal Limits on Massless High-Spin Particles}

\vspace{20pt}

{\large M. Porrati}

\vspace{12pt}

Center for Cosmology and Particle Physics\\
Department of Physics\\ New York University\\
4 Washington Pl.\\ New York, NY 10003, USA

\end{center}
\vspace{20pt}

\begin{abstract}
We present a model-independent argument showing 
that massless particles interacting with
gravity in a Minkowski background space can have at most spin two. This result
is proven by extending a famous theorem due to Weinberg and Witten to
theories that do not possess a gauge-invariant stress-energy tensor.
\end{abstract}

\end{titlepage}

\newpage

\section{Introduction}
The ever-so-many vacua of string theory~\cite{s} differ in almost any 
conceivable way from one another, yet they all have something in common:
none of them describes a 
four-dimensional Minkowski space with massless particles of spin larger 
than two. The graviton is always the highest-spin massless state. Massive
particles of any spin do exist, but they are always composite states, 
or unstable 
resonances. More precisely, their mass $M_{s>2}$ is always larger than
the string scale $M_S$, and they are not point-like: they possess form factors 
that give them a size $L \apprge 1/M_S \apprge 1/M_{s>2}$. 
Hadronic resonances also have a finite, nonzero size $L \sim 1/M$; 
classical spinning objects of course have $L \gg 1/M$. 

We take these facts as hints that high-spin particles become strongly 
interacting at a finite energy scale $\Lambda$. How is $\Lambda$ related to the
particle mass $M$ and its spin $s$ and how does the strong coupling regime 
manifests? In string 
theory we can give a concrete answer to this question: a
multitude of other states of spin less than $s$ exists at or below the mass 
$M$. Their multiplicity is exponential in $M$: $D(M)\sim \exp(cM/M_S)$. ($c$ is
a numerical constant that depends on the specific string theory being 
considered.) The limit $M_S\rightarrow 0$ is singular: it produces an infinite
number of massless states, whose interactions have not 
yet been properly understood.

No example exists of a string compactification with a light 
high-spin state, i.e. an
$s>2$ state with mass $M_{s>2} \ll M_S$. More generally, no theory is known 
where particles of spin $s>2$ and mass $M_{s>2}$ interact weakly up to an 
arbitrarily high energy scale. In all known examples, regardless of the 
precise functional relation 
linking $\Lambda$ to $M_{s>2}$, $\Lambda$  vanishes in the massless limit 
$M_{s>2}\rightarrow 0$. 

What we said applies to flat Minkowski backgrounds. In Anti de Sitter space, 
theories with infinitely many {\em massless} particles of arbitrary spin are
known~\cite{v}. Their spectrum cannot be consistently truncated to a finite
number of particles. Even more importantly, their interactions scale as 
inverse powers of the cosmological constant $\lambda$, so the flat space limit 
$\lambda \rightarrow 0$ is singular. In AdS the cosmological constant plays the
role of 
the mass  scale which determines the onset of the strong coupling regime.
Thus, far from being in contradiction with our previous flat-space examples, 
AdS models point out to the same conclusion: when the relevant mass scale 
of our theory goes to zero, be it $M$ or $\sqrt{|\lambda|}$, a high-spin 
particle becomes strongly interacting. In formulas, if we denote by 
$\tilde{M}$ the {\em largest} between $M$ and $\sqrt{|\lambda|}$,
the effective coupling of the theory at an energy scale $E$ behaves as 
\beq
g_{eff}(E) \sim \frac{E^{a+b}}{{M_*}^a \tilde{M}^b }, \qquad a,b>0.
\eeq{1}
To be general, we introduced another mass scale, $M_*$, 
which characterizes possible
interactions of the high-spin state with either itself or other particles.
Since all particles should interact at least with gravity, a universal choice
for $M_*$ is the Planck mass $M_{Pl}$.

Equation~(\ref{1}) is motivated by a simple observation: 
if a high-spin theory had
a well defined $M_{s>2}\rightarrow 0$ limit, then it would be possible to
construct massless high-spin theories interacting 
at least with gravity. Yet, strong
constraints exist in the literature, that forbid this possibility. All known
{\em no go} arguments or theorems have loopholes; aim of this paper is to
close some of those loopholes.

We will review the main existing {\em no go} theorems on interacting 
high-spin theories 
in Section 2; in particular, we will briefly re-derive the Weinberg-Witten 
theorem~\cite{ww}.
By a suitable weakening of its hypotheses, it will give us the desired 
{\em no go}\footnote{As for all {\em no go} theorems, ours should be rather 
called {\em don't go there}. These theorems often allow exceptions, and their 
constructive role is precisely to show which avenue one should not take in the 
search
for self-consistent theories.}, presented in Section 3. Specifically,
Section 3 presents our argument, showing that particles of spin larger 
than two 
cannot have gravitational interactions in Minkowski space. The proof parallels
the seminal results obtained in a Lagrangian framework by Aragone and Deser 
for spin 5/2~\cite{ad}; it extends their results beyond their local
field theory framework, and it generalizes it to arbitrary spins. 
Section 4 contains a discussion of our result, its limitations 
and possible extensions, as 
well as an application of the  methods of Section 3 
to the simpler case of charged particles in 
interaction with 
massless Abelian gauge fields. Section 4 also puts forward 
some speculations on 
how to circumvent our {\em no go} theorem.

\section{A Brief History of {\em No Go} Theorems}
An important obstruction to consistent interactions of high-spin massless 
particles was derived in 1964 by Weinberg~\cite{w64} using general properties
of the S-matrix. His result was extended to Fermions and specifically to
supersymmetric theories in~\cite{gp,gpv}. 

We shall review now Weinberg's result since we will use later one of its key 
techniques.
Consider an S-matrix element with $N$ external particles of four-momentum 
$p_i$, 
$i=1,..N$ and one massless spin $s$ particle of momentum $q$ and polarization 
vector $\epsilon^{\mu_1..\mu_s}(q)$. In the soft limit
$q\rightarrow 0$, it factorizes as (see fig. 1)
\beq
S(p_1,..,p_N,q,\epsilon)\approx \sum_{i=1}^N g^i{p^i_{\mu_1}...p^i_{\mu_s}
\epsilon^{\mu_1..\mu_s}(q) \over 2p^i q} S(p_1... p_N).
\eeq{2}
The polarization vector is transverse and traceless
\beq
q_\mu\epsilon^{\mu \mu_2..\mu_s}(q)=0, \qquad 
\epsilon_\mu^{\mu \mu_3..\mu_s}(q)=0.
\eeq{3}
\begin{picture}(100000,16500)
\drawline\fermion[\NW\REG](5000,10000)[5000]
\put(\pbackx,14000){$p_1$}
\put(7000,10500){$p_i+q$}
\drawline\fermion[\E\REG](5000,10000)[8000]
\put(\pbackx,\pbacky){$p_i$}
\drawline\gluon[\S\REG](12000,10000)[3]
\put(\pbackx,\pbacky){$q$}
\drawline\fermion[\SW\REG](5000,10000)[5000]
\put(\pbackx,6000){$p_N$}
\drawline\fermion[\N\REG](5000,10000)[5000]
\drawline\fermion[\NE\REG](5000,10000)[5000]
\drawline\fermion[\S\REG](5000,10000)[5000]
\drawline\fermion[\SE\REG](5000,10000)[5000]
\put(5000,10000){\circle*{20000}}
\put(18000,10000){$(g^ip^i_{\mu_1}...p^i_{\mu_s}\epsilon^{\mu_1..\mu_s}(q) /2p^i q)\times$}
\put(15500,10000){$\approx$}
\drawline\fermion[\NW\REG](35000,10000)[5000]
\put(\pbackx,14000){$p_1$}
\drawline\fermion[\E\REG](35000,10000)[5000]
\put(\pbackx,\pbacky){$p_i$}
\drawline\fermion[\SW\REG](35000,10000)[5000]
\put(\pbackx,6000){$p_N$}
\drawline\fermion[\N\REG](35000,10000)[5000]
\drawline\fermion[\NE\REG](35000,10000)[5000]
\drawline\fermion[\S\REG](35000,10000)[5000]
\drawline\fermion[\SE\REG](35000,10000)[5000]
\put(35000,10000){\circle*{20000}}
\label{fig1}
\end{picture}
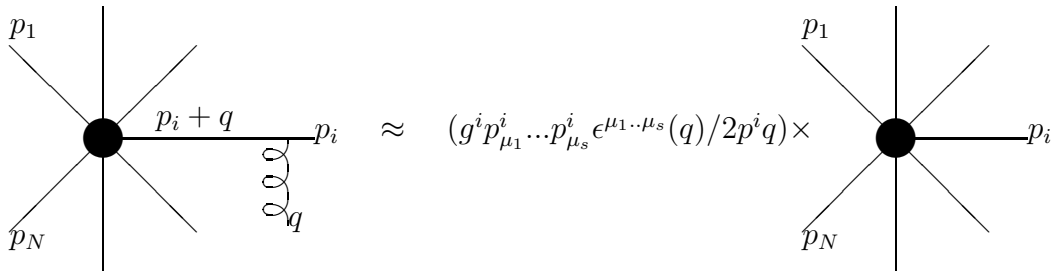
Figure 1: Factorization of S-matrix amplitude in the 
soft limit $q\rightarrow 0$.
\vskip .1in
\noindent
It gives a redundant 
description of the massless particle, which has only two physical 
polarizations. Redundancy is eliminated by demanding that the S-matrix is
independent of spurious polarizations
\beq
\epsilon_{spurious}^{\mu_1..\mu_s}(q)\equiv q^{(\mu_1}\eta^{\mu_2..\mu_s)}(q),
\qquad q_\mu\eta^{\mu\mu_1..\mu_{s-2}}(q)=\eta_\mu^{\mu\mu_1..\mu_{s-3}}(q)=0.
\eeq{4}
Factorization eq.~(\ref{2}) implies that spurious 
polarizations decouple only when
\beq
\sum_i g^i p^i_{\mu_1}...p^i_{\mu_{s-1}}=0, \qquad \forall p_i.
\eeq{5}
For generic momenta this equation has a solution only in two cases:
\begin{description}
\item{$s=1$} In this case eq.~(\ref{5}) reduces to $\sum_i g^i=0$, i.e. to 
conservation of charge.
\item{$s=2$} Eq.~(\ref{5}) becomes $\sum_i p_\mu^i=0$ and $g^i=\kappa$. 
The first equation enforces energy-momentum conservation, 
while the second gives the 
principle of equivalence: all particles must interact with the massless spin
two with equal strength $\kappa$.
\end{description}
For $s>2$ eq.~(\ref{5}) has no solution for generic momenta.

This argument shows that only scalars, vectors and spin two particles 
interacting at long distance as gravitons can give rise to 
long-distance interactions. The argument was extended  to Fermions 
in~\cite{gp,gpv}, where it was shown that interacting massless 
Fermions exist only
up to spin 3/2\footnote{Spin 3/2 Fermions were also shown to interact as the
supersymmetric partners of the graviton, i.e. the gravitini of 
supergravity theory.}.
Both~\cite{w64} and~\cite{gp,gpv} rely on the existence of processes 
in which the number
of spin $s$ particles changes by one unit. This is necessary to generate 
long-range interactions for integer $s$, but it leaves out the possibility of
interacting high-spin particles with a nonzero conserved charge. 
In particular, particles interacting only with the graviton according to the 
principle of equivalence are still allowed. Moreover, the interaction of
these particles could be softened by powers of $q_\mu$ in such a manner as to
cancel the offending pole in eq.~(\ref{2}). These particles do not generate
long range forces, but they can still interact. 

If we want to exclude completely high-spin massless particles, we must look
for a truly universal interaction, one that no particle can avoid. The best
choice is the gravitational interaction. Equation~(\ref{5}) shows that the
graviton interacts universally with matter in the soft limit $q\rightarrow 0$. 
Indeed, eq.~(\ref{5}) can be taken as the most general form of the equivalence
principle: all matter interacts with the graviton and in the limit 
$q\rightarrow 0$ the interaction vertex is 
$\kappa \langle f| T_{\mu\nu}|i\rangle $ ($|i\rangle, |f\rangle $ are the 
particle's initial and final states, respectively).  

Inconsistencies of gravitationally coupled high-spin massless particles were
specifically studied in~\cite{ad} for $s=5/2$. It is instructive to review the
argument presented there, since we will extend some of its techniques to 
a more general S-matrix framework in Section 4. 
Ref.~\cite{ad} writes down a local field theory for a spin $5/2$ field, 
described by a tensor-spinor $\psi_{ab}$, coupled to gravity, described by the
tetrad $e^\mu_a$. To quadratic order in $\psi_{ab}$ it reads 
\beq
S= \int d^4x e [ -{1\over 2} \bar{\psi}_{ab}\Dirac\psi_{ab} -\bar{\psi}_{ab}
\gamma_b \Dirac \gamma_c\psi_{ca} + 2\bar{\psi}_{ab}\gamma_bD_c\psi_{ca}
 +{1\over 4} \bar{\psi}_{aa}\Dirac \psi_{bb}
-\bar{\psi}_{aa}D_b\gamma_c\psi_{bc} ].
\eeq{6}
The field $\psi_{ab}$ gives a redundant description of the spin 5/2 state. In 
the free theory this redundancy is eliminated by the gauge invariance 
\beq
\delta\psi_{\mu\nu}=\partial_\mu\epsilon_\nu + \partial_\nu\epsilon_\mu,\;\;\;
\gamma^\mu\epsilon_\mu=0.
\eeq{7}
Upon covariantization, derivatives are replaced by covariant derivatives, but
the gauge transformation is otherwise unaffected
\beq 
\delta\psi_{ab}=D_a\epsilon_b + D_b\epsilon_a,\;\;\;
\gamma^a\epsilon_a=0.
\eeq{8}
Under the gauge transformation~(\ref{8}) action~(\ref{6}) transforms as
\beq
\delta
S=-4\int d^4 x e \bar{\epsilon}_a\gamma_b\psi_{cd}R^{abcd}.
\eeq{9}
So, the action is only invariant in flat space $R^{abcd}=0$; in other words,
gauge modes decouple only in the free theory. 
It is quite immediate to convince oneself that this inconsistency cannot be
cured by adding non-minimal terms to action~(\ref{6}) that are both {\em local
and regular in the neighborhood of flat space}. If the  last 
requirement is removed, as it is possible in a theory that makes sense
e.g. in Anti de Sitter space, but which does not allow for a flat space limit,
then we can not only decouple the gauge modes, 
but also write down a consistent 
theory of high-spin massless Fermions. To achieve consistency beyond the lowest
perturbative order, one must nevertheless introduce (infinitely many) new
massless states besides the spin 5/2 one\footnote{Literature on this subject 
is vast and complex. Comprehensive reviews of (Bosonic) high-spin theories
in AdS, with extended bibliography, can be
found in~\cite{v}.}.

The cosmological constant $\lambda <0$ appearing in AdS space defines a mass
parameter $O(\sqrt{|\lambda|})$. Massive spin 5/2 also evades rather trivially 
the no-go, since the gauge invariance is already broken by the mass term 
present in the massive analog of eq.~(\ref{6}). Interacting, massive high-spin 
theories are 
not algebraically inconsistent, but they do manifest pathologies ranging from 
superluminal propagation and ghosts in external coherent fields~\cite{vz} to
strong coupling behavior at a finite energy scale~\cite{p,cpd,pr}.
In fact, in AdS space one should expect no operational difference between 
massless particles and particles with Compton wavelength 
$\lambda_{Compton}\sim 1/m $ larger than the AdS curvature radius 
$R_{AdS}\sim 1/\lambda$. This expectation is 
confirmed by the absence of mass discontinuities in their 
propagators~\cite{p2,gvn,dw,fma}.

Returning now to massless particles in asymptotically Minkowski space, we could
imagine that the inconsistency evidenced by eq.~(\ref{9}) is due to the most
important
implicit assumption inherent to the Lagrangian formalism: locality. 
Eqs.~(\ref{6}) and ff assume a standard kinetic term for $\psi_{ab}$, and
local interactions. Could it
be that a carefully chosen form factor for gravitational interactions, 
tantamount to some reasonable non-locality in the Lagrangian can cure the 
problem? How does our result depend on the field representation 
chosen for the spin 5/2 particle? Could a non-minimal description, involving
a {\em larger} gauge invariance than that in eqs.~(\ref{7},\ref{8}) be 
consistent 
after all? To answer all these questions, we must use a truly universal 
formalism, dealing with matrix elements of observables. In other words, we 
should go back to the S-matrix language and use it to analyze an unavoidable
interaction: scattering of massless particles off soft gravitons. 

This analysis was done in~\cite{ww}, where a particular matrix element was
considered: elastic scattering of a spin $s$  massless particle off a single
soft graviton. The initial and final polarizations of the spin $s$ particle are
identical, say $+s$, its initial momentum is $p$ and its final momentum is 
$p+q$. The graviton is off-shell with momentum $q$.
The matrix element is
\beq
\langle +s,p+q| T_{\mu\nu} | +s,p\rangle .
\eeq{10}
In the soft limit $q\rightarrow 0$ the matrix element is completely
determined by the equivalence principle. 
Using the relativistic normalization for one-particle states, 
$\langle p| p'\rangle =  2p_0(2\pi)^3 \delta^3({\bf p} -{\bf p'})$, we get
\beq
\lim_{q\rightarrow 0} \langle +s,p+q| T_{\mu\nu} | +s,p\rangle =p_\mu p_\nu.
\eeq{11}

Since $q$ is space-like, there exists a frame --the ``brick wall'' frame-- 
in which 
\beq
p^\mu= (|{\bf q}|/2, {\bf q}/2),\qquad q^\mu= (0 , -{\bf q}),
\qquad p^\mu+q^\mu= (|{\bf q}|/2, -{\bf q}/2).
\eeq{12}
A rotation $R(\theta)$ by an  angle $\theta$ around the ${\bf q}$ direction
acts on the one-particle states as
\beq
R(\theta)|p,+s\rangle = \exp(\pm i\theta s)|p,+s\rangle, \qquad
R(\theta)|p+q,+s\rangle = \exp(\mp i\theta s)|p+q,+s\rangle,
\eeq{13}
since $R(\theta)$ is a rotation of $\theta$ around ${\bf p}$ but of $-\theta$
around ${\bf p} + {\bf q} = -{\bf p}$. Under space rotations, $T_{\mu\nu}$ 
decomposes into two real scalars, one vector and one symmetric traceless 
tensor. In the standard basis where the commuting variables 
are the total angular momentum 
and its projection along the axis ${\bf q}$, 
these field are represented by spherical tensors: 
$T_{0,0}$, $T_{1,m}$, $m=0,\pm 1$
and $T_{2,m}$, $m=0,\pm1,\pm2$. Here we have combined the two 
real scalars into a complex 
scalar. In this basis one gets the trivial identity
\beq
e^{\pm 2i\theta s} \langle +s,p+q| T_{j,m} | +s,p\rangle =
\langle +s,p+q| R^\dagger T_{j,m} R | +s,p\rangle = e^{i\theta m}
\langle +s,p+q| T_{j,m} | +s,p\rangle.
\eeq{14}
For $s>1$, the only solution to this equation is 
$\langle +s,p+q| T_{\mu\nu} | +s,p\rangle=0$. 

{\em If $T_{\mu\nu}$ is a tensor under Lorentz transformations} then
eq.~(\ref{14}) implies that the matrix element~(\ref{10}) vanishes in all
frames, in contradiction with the equivalence principle eq.~(\ref{11})!

The crucial assumption here is that $T_{\mu\nu}$ (better, its matrix element
between massless spin $s$ states) is a Lorentz tensor. The assumption is far 
from innocuous. In particular, neither the gravitino (spin 3/2) nor the 
graviton (spin 2) satisfy this hypothesis~\cite{ww}. This happens because
both spin 3/2 and spin 2
have gauge invariances (local supersymmetry and diffeomorphisms, respectively)
and their stress-energy tensor is not gauge invariant. The stress-energy tensor
derived from Lagrangian higher-spin theories exhibits the same phenomenon: to
achieve gauge invariance one must forgo manifest Lorentz covariance~\cite{dw2}.

In fact, non-Lorentz covariance of gauge non-invariant operators is a
familiar fact in field theory. The simplest example is offered by the matrix
element of the EM gauge potential $A_\mu$  in between the vacuum and a
one-photon state, $\langle 0 | A_\mu | s,p\rangle$ ($s=\pm 1 $). One can 
always choose a complete set of polarization vectors for $A_\mu$ such that
$\langle 0 | A_0 | s,p\rangle =0$. A Lorentz boost $L$ leaves the vacuum
invariant and transforms the one particle state as 
$L  | s,p\rangle = \exp[i\theta(L,p)s]  | s,p\rangle $; therefore, 
$\langle 0 | L^\dagger A_0 L | s,p\rangle =
\exp[i\theta(L,p)s] \langle 0 | A_0 | s,p\rangle =0 $ in all frames. This is of
course incompatible with the transformation law of a Lorentz vector. Indeed
a Lorentz boost transforms inhomogeneously the matrix element. In 
infinitesimal form, the transformation law is a standard Lorentz transformation
plus a compensating gauge transformation
\beq
\delta_\omega \langle 0 | A_\mu | s,p\rangle = 
\omega_\mu^\nu \langle 0 | A_\nu | s,p\rangle + p_\mu \Phi(p,\omega).
\eeq{15}
One {\em can} define matrix elements for $A_\mu$ that transform as Lorentz 
vectors, but only at the price of introducing non-physical states which
correspond to spurious polarizations.
Likewise, one can define $T_{\mu\nu}$ matrix elements that transform as 
Lorentz tensors by introducing spurious polarizations. These facts
are the crucial ingredient in our treatment of high-spin massless
fields. 
\section{An Improved {\em No Go} Theorem}
To replace eq.~(\ref{10}) with a Lorentz covariant matrix element 
we need extra spurious states, besides the physical ones given in 
eq.~(\ref{13}). These states mix with physical polarizations under Lorentz 
transformations: $v_{phys} --> v_{phys} + v_s$\footnote{We saw this phenomenon 
at work in the case of spin one in eq.~(\ref{15}).}. Physical and spurious
states together thus span a reducible but not block-diagonal
representation of the Lorentz group. 
Spurious states must decouple from all physical matrix elements and in
particular from S-matrix amplitudes. If we denote with $v$ all one-particle, 
spin $s$ states, whether or not spurious, the matrix element is
$\langle v', p+q | T_{\mu\nu} | v, p \rangle $.  It is not an S-matrix element
yet, since the graviton (and {\em only} the graviton) is off-shell. A 
convenient method to derive the 
S-matrix is to perform the standard perturbative expansion of the effective
action
\beq
A={1\over 16\pi G} \int d^4 x \sqrt{-g}R(g) + {1\over 2} 
\int {d^4 q\over (2\pi)^4} h^*_{\mu\nu} (q) [
\langle v', p+q | T^{\mu\nu} | v, p \rangle + {\cal T}^{\mu\nu} ] + 
{\cal O} (h^2).
\eeq{16}
The standard Einstein action with Newton's constant $G$ and metric 
$g_{\mu\nu}$ has been supplemented here with 
certain interaction terms, written in a perturbative expansion around
flat space ($ g_{\mu\nu}=\eta_{\mu\nu} + h_{\mu\nu} $). The linear 
interaction terms include our matrix element and another effective stress
energy tensor ${\cal T}_{\mu\nu}$, which summarizes the effect of any other
matter field. 
Eq.~(\ref{16}) is not only a convenient bookkeeping device, but it also
gives us the most general condition for the decoupling of a spurious 
polarization
$v_{s}$. Decoupling occurs when one can reabsorb the change in the matrix 
element due to the
substitution $v \rightarrow v + v_s$ with a {\em local}
field redefinition of the graviton field $h_{\mu\nu}$. This happens because
the S-matrix is
independent of such redefinition~\cite{sal}. To linear order in $h_{\mu\nu}$,
Einstein's equations become 
\bea
L_{\mu\nu}^{\;\;\;\;\rho\sigma}h_{\rho\sigma}(q)&=& 16\pi G[
\langle v', p+q | T_{\mu\nu} | v, p \rangle + {\cal T}_{\mu\nu}], \nonumber\\
L_{\mu\nu}^{\;\;\;\;\rho\sigma} &=& \delta^{\rho\sigma}_{\mu\nu} q^2 
-\delta_{\mu\nu}\delta^{\rho\sigma}q^2 -\delta_\mu^\rho q_\nu q^\sigma 
-\delta_\nu^\rho q_\mu q^\sigma + \delta^{\rho\sigma} q_\mu q_\nu  
+ \delta_{\mu\nu} q^\rho q^\sigma .
\eea{17}
To the same order, we then get a necessary condition for the consistency of 
gravitational  interactions of high-spin massless particles:
\beq
\langle v, p+q | T_{\mu\nu} | v_s, p \rangle = 
L_{\mu\nu}^{\;\;\;\;\rho\sigma}\Delta_{\rho\sigma}(q),
\eeq{18}
with $\Delta_{\mu\nu}(q)$ analytic in a neighborhood of $q=0$.
The required field redefinition is 
$h_{\mu\nu} \rightarrow h_{\mu\nu} + 16\pi \Delta_{\mu\nu}(q)$.

Eq.~(\ref{18}) weakens the hypotheses of the Weinberg-Witten theorem by 
allowing the matrix element to depend non-trivially on spurious polarizations. 
In Lagrangian language, this means that the stress-energy is not separately 
gauge invariant, though the action is.
The Weinberg-Witten result is recovered by demanding the stronger condition
$\Delta_{\mu\nu}(q)=0$, i.e. gauge invariance.

Eq.~(\ref{18}) does not guarantee the existence of a consistent theory, since
inconsistencies can show up in contact terms at ${\cal O}(h^2)$, but if not
satisfied it signals a lethal inconsistency, since in that case no amount of
extra fields or extra interactions can cancel the $v_s$ dependent change in the
action.
Notice that while $\Delta_{\mu\nu}(q)$ must be analytic in $q$ 
for small $q$, no such
requirement holds for the matrix element itself. This is a first advantage of
the S-matrix formalism over the Lagrangian analysis of ref.~\cite{ad}, which we
summarized in Section 2. In a Lagrangian framework one must necessarily 
assume locality of the
matrix element itself; moreover, one is still left with the doubt that a field
redefinition of $\psi_{\mu\nu}$ may change the analysis. In our case, since
the initial and final spin $s$ states are on-shell, 
no such redefinition exists.

The last observation also answers another question about the generality of our 
result: can it depend on the particular choice of spurious states we are
going to make? No, it can't. A non-minimal choice of spurious states 
means to introduce a larger set of them, which we can denote with $\{V_s\}$. 
By setting some of them to zero, we go back to our minimal choice 
(to be defined shortly), $\{ v_s \} \subset \{ V_s\}$. 
Independence of $\{V_s\}$ thus implies independence of all $v_s$, which implies
eq.~(\ref{18}).
\subsection{Fermions}
The matrix element $\langle v', p+q | T_{\mu\nu} | v, p \rangle $ is bilinear
in $v,v'$ and it otherwise depends only on the momenta. For spin $s$, 
the minimum set of
spurious states needed to write a nonzero conserved, symmetric tensor is
given by Dirac spinor-tensors $v_{\alpha,\; \mu_1...\mu_n}(p)$, $s=n+1/2$. 
They are symmetric in the vector indices $\mu_1,,\mu_n$ and satisfy the
constraints  
\beq
/\!\!\!p v_{\mu_1,..\mu_n}(p)=0, \qquad p^{\mu_1} v_{\mu_1,..\mu_n}(p)=0,
\qquad \gamma^{\mu_1} v_{\mu_1,..\mu_n}(p).
\eeq{19}
We are interested in initial and final states with the same physical helicity 
$+s$, so on the representatives of the initial state ($u$) and final state 
($v$) we impose 
\beq
\gamma^5 u_{\mu_1,..\mu_n}(p) =u_{\mu_1,..\mu_n}(p), \qquad
\gamma^5 v_{\mu_1,..\mu_n}(p+q) =v_{\mu_1,..\mu_n}(p+q).
\eeq{20}

In the kinematical configuration of interest, there exist two independent 
light-like vectors: $p$ and $p+q$. The space-like vector $q$ can be used to
define $n+1$ algebraically independent spinor-tensors
\beq
u^k_{\mu_1,..,\mu_k}(p)\equiv q^{\mu_{k+1}}...q^{\mu_n} u_{\mu_1,..\mu_n}(p),
\qquad k=0,..,n.
\eeq{21}
Their algebraic independence is verified by writing down their explicit 
parametrization in the brick wall frame eq.~(\ref{12}). Introduce first of all
vector polarizations $\epsilon$ and an on-shell spinor $\chi$
\beq
\epsilon^3_\mu=(-1,1,0,0),\qquad \epsilon^\pm_\mu=(0,0,1,\pm i),\qquad
\gamma^5 \chi=\chi,\qquad (\gamma^0 - \gamma^1) \chi =0.
\eeq{22}
The last equation is the on-shell condition $/\!\!\! p \chi(p)=0$; the last
two conditions imply $(\gamma^2 + i \gamma^3) \chi =0$. The $n+1$ 
spinor tensors
\beq
u_{\mu_1,..\mu_n}^{(k)} \equiv \epsilon^+_{(\mu_1} .... \epsilon^+_{\mu_k} 
\epsilon^3_{\mu_{k+1}}\epsilon^3_{\mu_n)}\chi, \qquad k=0,..,n,
\eeq{23}
are evidently linearly independent, they satisfy the 
constraints~(\ref{19},\ref{20}) and obey
\beq
u_{\mu_1,..\mu_l}^{(k)l} \equiv 
q^{\mu_{l+1}}...q^{\mu_n} u_{\mu_1,..\mu_n}^{(k)} \left\{\begin{array}{l}
=0, \qquad \mbox{for }l<k, \\ \neq 0, \qquad \mbox{for }l\geq k .
\end{array}\right. 
\eeq{24}
The triangular linear system~(\ref{24}) defines $n+1$ independent spinor 
tensors. Eq.~(\ref{21}) or~(\ref{24}) parametrize one physical polarization of
helicity $s=n+1/2$ and $n$ spurious polarizations with 
$s=1/2,..., n-1/2$\footnote{$s<0$ states are eliminated by the chirality
projection $\gamma^5 u_{\mu_1,..\mu_n} = u_{\mu_1,..\mu_n}$.}.

Constraints~(\ref{19},\ref{20}) and the on-shell condition on momenta, $p^2=
(p+q)^2=0$, vastly reduce the possible terms in the 
matrix element of interest. 
A short reflection suffices to convince oneself that its most general form is
\beq
\langle v, p+q | T_{\mu\nu} | u, p \rangle = \sum_{k=0}^n A^k \bar{v}^k
(p+\alpha^k q)_{(\mu} \gamma_{\nu)} u^k + \sum_{k=1}^n B^k \bar{v}^k_{(\mu}
\gamma^{}_{\nu)} u^{k-1} + \sum_{k=1}^n C^k \bar{v}^{k-1}_{(\mu}
\gamma^{}_{\nu)} u^{k}.
\eeq{25}
The coefficients $A^k$ ,$B^k$, $C^k$ and $\alpha^k$ are functions of $q^2$ 
which in principle can be singular at $q^2=0$. A first constraint on the
singularity is due to the principle of equivalence that demands
\beq
 \lim_{q\rightarrow 0} \langle v, p+q | T_{\mu\nu} | u, p \rangle =
p_\mu p_\nu.
\eeq{26}
This equation implies
\bea
\lim_{q\rightarrow 0} A^n(q) &= & 1,\label{27} \\
\lim_{q\rightarrow 0} A^{k}(q)q^{2(n-k)}&=& 0, \qquad k<n ,\label{28}\\
\lim_{q\rightarrow 0} \alpha^k(q) A^k(q) q^{2(n-k)-1} &=& 0 , \label{29} \\
\lim_{q\rightarrow 0} B^k(q) q^{2(n-k)+1}&=&0 , \label{30}\\
\lim_{q\rightarrow 0} C^k(q) q^{2(n-k)+1}&=& 0 . 
\eea{31}
Conservation of $T_{\mu\nu}$ implies that the matrix element~(\ref{25}) is
divergenceless 
\beq
q^\mu \langle v, p+q | T_{\mu\nu} | u, p \rangle =0. 
\eeq{31a}
This
yields the further constraints
\bea
A^k (\alpha^k -1/2) q^2 + B^{k+1} + C^{k+1} &=& 0, \qquad k=0,.., n-1 ,
\label{32} \\
\lim_{q\rightarrow 0}\alpha^n (q) &=&1/2, \qquad 
\lim_{q\rightarrow 0} A^n(q) =1.
\eea{33}
Though not strictly necessary to prove our result, eq.~(\ref{33}) is useful
since it simplifies the structure of the matrix element. In particular, 
together with the mass-shell conditions~(\ref{19}) it makes the matrix element 
transverse and traceless. 

In reality, constraints~(\ref{27}-\ref{31}) are too weak, because if any of 
the  coefficients 
$A^k$ ,$B^k$, $C^k$ and $\alpha^k A^k$ had a singularity $1/q^2$~\footnote{For
instance $A^k(q)=A_r^k(q)q^{-2}$, $A^k_r(q)=$ regular and nonzero at $q^2=0$.}
then vertex~(\ref{25}) would imply the existence of another massless spin 2
field (it couples to a transverse-traceless vertex!) which mixes linearly with
the graviton. This linear mixing contradicts Weinberg's 
uniqueness theorems for soft gravitons~\cite{w65}. 
It also violates the principle of equivalence --which we
assumed (and need) to prove or theorem-- either because it
implies the existence of a second massless graviton that couples only to some
type of matter (massless high-spin) or because it re-sums to give the 
graviton a
mass.  A singularity stronger than $1/q^2$ is even worse since it implies the
existence of a spin two ghost mixing linearly with the ordinary graviton 
(see fig. 2).
\begin{picture}(100000,16500)
\drawline\fermion[\NW\REG](15000,10000)[5000]
\put(\pbackx,14000){$v$}
\put(18500,11000){$q$}
\put(18000,8000){$1/q^2$}
\drawline\photon[\E\REG](15000,10000)[6]
\put(\pbackx,\pbacky){\circle*{500}}
\drawline\gluon[\E\REG](\pbackx,\pbacky)[6]
\put(22000,8000){graviton}
\drawline\fermion[\SW\REG](15000,10000)[5000]
\put(\pbackx,5500){$u$}
\put(15000,10000){\circle*{20000}}
\label{fig2}
\end{picture}
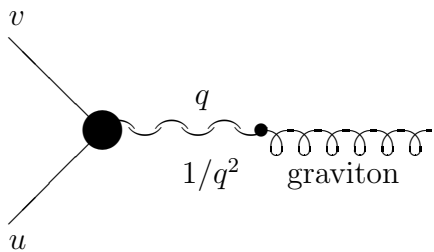
Figure 2: A singular vertex implies the existence of an additional massless
particle mixing with the graviton.
\vskip .1in
\noindent

We have introduced extra polarizations to ensure that the $T_{\mu\nu}$ matrix
element transforms covariantly. Now we must check under which conditions 
spurious polarizations do decouple. Spurious states have the form
\beq
u_{s\;\mu_1...\mu_n}(p)=p^{}_{(\mu_1}\epsilon_{\mu_2.... \mu_n)},
\eeq{34}
where $\epsilon_{\mu_1.... \mu_{n-1}}$ is on shell, transverse and 
gamma-transverse. For the spurious state~(\ref{34}), 
the spinor-tensors given in eq.~(\ref{21}) have the form
\beq
u^{k}_{s\;\mu_1...\mu_k}(p)= p_{(\mu_1}\epsilon^{k-1}_{\mu_2.... \mu_k)} 
-(n-k){q^2 \over 2} \epsilon^{k}_{\mu_1.... \mu_k} ,
\qquad 
\epsilon^{k}_{\mu_1.... \mu_k} \equiv q^{\mu_{k+1}}...q^{\mu_n}
\epsilon^{k}_{\mu_1.... \mu_n}.
\eeq{35}
Matrix element~(\ref{25}) is transverse and traceless, therefore the decoupling
condition~(\ref{18}) simplifies to
\beq
\langle v, p+q | T_{\mu\nu} | u_s, p \rangle = q^2 \Delta_{\rho\sigma}(q).
\eeq{36}
Substitution of eqs.~(\ref{34},\ref{35}) into eq.~(\ref{25}) then
yields a set of 
recursion relations among the coefficients $A^k,...,C^k$:
\bea
-kA^k -{q^2\over 2}(n+1-k)A^{k-1} +C^k &=& {\cal O}(q^2), \qquad
k=1,...,n;\label{37}\\
-k\alpha^k A^k -{q^2\over 2}(n+1-k)\alpha^k A^{k-1} &=& {\cal O}(q^2), 
\qquad k=1,...,n;
\label{38}\\
-(k-1)B^k -{q^2\over 2}(k+2-k)B^{k-1} &=& {\cal O}(q^2), \qquad k=2,...,n;
\label{39}\\
-(k-1)C^k -{q^2\over 2}(k+1-k)C^{k-1} &=& {\cal O}(q^2), \qquad k=2,...,n.
\eea{40}
As we have seen earlier, no coefficient in eq.~(\ref{25}) can be more 
singular than $1/q^2$. So in particular
\beq
\lim_{q\rightarrow 0}q^2 C^1(q)=\lim_{q\rightarrow 0}q^2 A^0(q)=0.
\eeq{41}
Recursion relations~(\ref{37}, \ref{40}) then imply
\beq
\lim_{q\rightarrow 0}A^n(q)=0, \qquad n>1,
\eeq{42}
in contradiction with eq.~(\ref{27}), $A^n(0)=1$, which is nothing else than 
the equivalence principle! 

This completes our proof: only when spurious polarizations decouple from the
cubic vertex~(\ref{25}) a chance exists for massless high-spin fields to
interact with gravity, but decoupling contradicts the universality of
gravitational interactions! 

Our argument rules out interactions for Fermions of spin $s> 3/2$. It still
allows for gravitational interactions of  spin 3/2 particles. This is not
surprising since supergravity theories provide many examples of massless
spin 3/2 particles consistently interacting with gravity and other fields.

Notice that our argument does not rule out exotic high-spin interacting 
theories, but it shows that these theories do not have any common interaction
with physical matter, which must interact with gravity universally.

Notice too that our argument relies crucially on the exact masslessness of the
graviton. In the conclusions, we will briefly discuss the dynamics of
high-spin massless particles in theories where gravity changes in the 
infrared, as in massive gravity or in the DGP model~\cite{dgp}.
\subsection{Bosons}
The proof of our theorem in the Bosonic case parallels that we gave for Fermions. Polarizations 
are now described by the on-shell, symmetric, transverse and traceless tensors
\beq
U_{\mu_1..\mu_s}(p), \qquad p^{\mu_1} U_{\mu_1..\mu_s}(p)=
U^{\mu_2}_{\mu_2..\mu_s}(p)=0.
\eeq{43}
Spurious polarizations read
\beq
U_{\mu_1..\mu_s}(p) = p_{(\mu_1}\epsilon_{\mu_2...\mu_s)}, \qquad
p^{\mu_2}\epsilon_{\mu_2...\mu_s}=
\epsilon^{\mu_3}_{\mu_3...\mu_s}=0.
\eeq{44}
In complete analogy with the Fermion treatment, we use contraction with
$q^\mu$ to define 
\beq
U^k_{\mu_1..\mu_k}(p)=q^{\mu_{k+1}}...q^{\mu_s}U_{\mu_1..\mu_s}(p),\qquad
\epsilon^k_{\mu_1...\mu_k}=q^{\mu_{k+1}}...q^{\mu_{s-1}}
\epsilon_{\mu_1...\mu_{s-1}}.
\eeq{45}
Contraction of the spurious polarizations defined by eq.~(\ref{44}) results in
\beq
U^k_{\mu_1..\mu_k}(p)=k p_{(\mu_1}\epsilon^{k-1}_{\mu_2...\mu_k)}-
(s-k){q^2\over 2} \epsilon^k_{\mu_1...\mu_k}.
\eeq{46}

The most general form of the matrix element 
$\langle V, p+q | T_{\mu\nu} | U, p \rangle $ is now
\bea
\langle V, p+q | T_{\mu\nu} | U, p \rangle &=& \sum_{k=0}^s {A^k \over 2}
(p_\mu p_\nu + 2\alpha^k p_\mu q_\nu + \tilde{\alpha}^k q_\mu q_\nu +
\hat{\alpha}^k \eta_{\mu\nu}) \bar{V}^k
U^k + \nonumber \\
&& \sum_{k=0}^{s-1} B^k (p_\nu + \beta^k q_\nu)\bar{V}^{k+1}_\mu U^k + 
\sum_{k=0}^{s-1} C^k (p_\nu + \gamma^k q_\nu)\bar{V}^k U^{k+1}_\mu +
\nonumber \\
&& 
\sum_{k=0}^{s-1} D^k \bar{V}^{k+1}_\mu U^{k+1}_\nu + 
\sum_{k=0}^{s-2} E^k \bar{V}^{k+2}_{\mu\nu} U^k + 
\sum_{k=0}^{s-2} F^k \bar{V}^k U^{k+2}_{\mu\nu}.
\eea{47}
By contracting expansion~(\ref{47}) with $q^\mu$ and equating all 
algebraically independent terms to zero, we enforce conservation of the
stress-energy tensor. As for Fermions, while not necessary to prove our result,
$q^\mu$ transversality somewhat simplifies the algebra. Specifically, 
vanishing of terms proportional to $\bar{V}^{k+1}_\mu U^k$ yields the equation
\beq
B^k(\beta^k-1/2)q^2 +D^k +E^k=0,\qquad k\leq s-1,
\eeq{48}
where we defined $E^{s-1}\equiv 0$.
Setting to zero terms proportional to $\bar{V}^kU^{k+1}_\mu $ we get
\beq
C^k(\gamma^k -1/2)+D^k + F^k=0, \qquad k\leq s-1,\qquad F^{s-1}\equiv 0.
\eeq{49}
Finally, vanishing of terms proportional to $p_\mu \bar{V}^k U^k$ 
implies
\beq
A^s(\alpha^s-1/2)=0, \qquad A^k(\alpha^k-1/2)q^2 + B^k + C^k=0, 
\quad k\leq s-1,
\eeq{50}
while vanishing of terms proportional to $q_\mu \bar{V}^k U^k$ results in
\beq
A^s[\hat{\alpha}^s+q^2(\tilde{\alpha}^s-\alpha^s/2)]=0, \qquad 
A^k[\hat{\alpha}^k+q^2(\tilde{\alpha}^k-\alpha^k/2)] + 
B^k\beta^k + C^k\gamma^k=0, 
\quad k\leq s-1.
\eeq{51}
Matrix element~(\ref{47}) is traceless for
\beq
A^0\hat{\alpha}^0=B^0/4, \qquad A^s\hat{\alpha}^s=-D^{s-1}/4, \qquad 
A^k\hat{\alpha}^k=(B^k-D^{k-1})/4, \qquad 
k=1,...,s-1. 
\eeq{52}
A generic conserved symmetric tensor can be decomposed into a 
transverse-traceless ($TT$) part and a scalar remnant as $\Theta_{\mu\nu}=
\Theta^{TT}_{\mu\nu} + (q_\mu q_\nu-q^2\eta_{\mu\nu})\Theta^S $. Of course, if
spurious polarizations decouple, they do so separately in the $TT$ and $S$ 
parts of matrix element~(\ref{47}); therefore, we can assume as well that it 
is traceless. In this case, the most general condition for decoupling is
eq.~(\ref{36}), again with $\Delta_{\mu\nu}(q)$ analytic at $q=0$.

Substituting the spurious polarizations~(\ref{46}) into eq.~(\ref{47}) 
and equating all algebraically independent terms in the latter to 
$q^2\Delta_{\mu\nu}(q)$, we get 
several constraints on the small-$q$ behavior of the coefficients 
$A^k,..,F^k$ and $\alpha^k,..,\gamma^k$. In particular, terms proportional to 
$p_\mu p_\nu \bar{V}^k \epsilon^k$ give the condition
\beq
C^k -(k+1)A^{k+1} -(s-k){q^2\over 2} A^k ={\cal O}(q^2), \qquad k=0,..,s-1.
\eeq{53}
Terms proportional to $p_{(\mu} \bar{V}^{k-1}\epsilon^k_{\nu)}$ give
\beq
(k+1)C^{k+1} -(s-k-1){q^2\over 2} C^k +E^k= {\cal O}(q^2), \qquad k=0,...,s-2.
\eeq{54}
Finally, terms proportional to $\bar{V}^{k+1}_{\mu\nu}\epsilon^{k-1}$ give
\beq
(k+1)E^{k+1} + (s-k) {q^2\over 2} E^k = {\cal O}(q^2), \qquad k=0,...,s-3.
\eeq{55}
Notice that we need $s\geq 3$ to obtain this full set of equations. 

As in the Fermionic case, the coefficients $A^k,..,F^k$ can be singular in the 
$q\rightarrow 0$ limit, but 
they must diverge less than $1/q^2$. In this case eq.~(\ref{55}) implies 
\beq
\lim_{q\rightarrow 0} E^{s-2}(q)=0.
\eeq{56}
The vanishing of $E^{s-2}$ in the soft limit $q\rightarrow 0$ and  
eq.~(\ref{54}) then imply
\beq
\lim_{q\rightarrow 0} C^{s-1}(q)=0.
\eeq{58}
Substituting this last equation into~(\ref{53}) we arrive at the main 
result of this subsection:
\beq
\lim_{q\rightarrow 0} A^{s}(q)=0.
\eeq{59}
The vanishing of $A^s$ at zero graviton momentum is in contradiction with the 
equivalence principle, which 
demands $\lim_{q\rightarrow 0} A^{s}(q)=1$.
So, massless
Bosons of spin $s\geq 3$ cannot couple with gravity. It is straightforward 
to check that the set of equations~(\ref{47}-\ref{54}) has a  solution 
satisfying the correct soft limit dictated by the principle of equivalence
for $s=2$. This is possible thanks to the fact that for $s=2$ there
is one less constraint to satisfy, namely eq.~(\ref{55}).

\section{Discussion and Conclusions}
In this paper, we borrowed ideas from the Weinberg-Witten {\em no go} 
theorem~\cite{ww} as well as from known
results on inconsistencies of gravitational coupling of 
high-spin massless particles, 
specifically from
ref.~\cite{ad}, to show that no massless high-spin particle can be 
consistently coupled to gravity in flat 
space. The theorem exploited a particular one-graviton matrix element, 
whose form is constrained in the 
soft-graviton limit by the equivalence principle. 
We showed that, under fairly general assumptions, this constraint is 
incompatible with the decoupling of the spurious polarizations 
that one must necessarily introduce 
to write down the matrix element in a Lorentz covariant form. 

The proof of the theorem was straightforward but not stunningly elegant. 
Clumsiness was the price we paid to 
allow for some mild non-locality in the matrix element. 
In particular, we did {\em not} demand  analyticity at  $q^2=0$ 
for the coefficients in
our matrix-element expansion eqs.~(\ref{25}) or~(\ref{47}). 
Had we done so, we could have extended the
matrix element to complex values of the momenta and 
put the graviton too on-shell, since the condition
$p^2=q^2=(p+q)^2=0$ does have nontrivial complex solutions. 

\subsubsection*{{\em No-Go} in the BCFW Construction}
Analyticity for complex momenta is one of the main ingredients 
in the BCFW construction
of S-matrix tree-level amplitudes~\cite{bcfw}. The use of complex 
momenta\footnote{Or equivalently the use of a space-time(s) 
metric of signature $(2,2)$.}
not only allows us to write non-vanishing three-particle 
on-shell vertices, but it also 
allows us to deform two of the momenta in an arbitrary scattering 
amplitude along a special
complex direction according to the formula
\beq
p_1 \rightarrow p_1 + z q, \qquad p_2 \rightarrow p_2 - z q, 
\qquad p_i^2=p_iq=q^2=0,\qquad i=1,2.
\eeq{57}
Any tree-level amplitude now becomes a rational function of the complex 
parameter $z$, with 
at most simple poles~\cite{bcfw}. So, if a particular 
amplitude vanishes at large $z$, then it can be computed by 
knowing the position of the poles
and the value of the residues. These are {\em on-shell} data that 
are completely 
specified by the three-point on-shell vertices. 
By applying the BCFW construction
to a four-particle amplitude involving the exchange of a graviton, 
Benincasa and 
Cachazo~\cite{bc} proved in an elegant manner that the only massless particles 
of spin $s>1$ that can be
coupled to gravity are the graviton and the gravitino~\cite{bc}, 
and that they interact 
exactly as in supergravity. The most restrictive assumption in
their construction is precisely the
vanishing of the amplitude at large $z$. This property is far from obvious. 
It requires extra assumptions on the theory under consideration, in addition
to Lorentz and gauge (or diffeomorphism) invariance~\cite{a-hk}.
We chose instead to keep our argument general
even at the price of weakening our result. 
\subsubsection*{}
One notable weakness of our argument is that it does {\em not} 
forbid the existence of more than one graviton; convincing arguments against 
this possibility have been given in the literature~\cite{w65,bdgh}. 
Its main strength is that it does not rely on a particular field 
parametrization or on assuming a specific Lagrangian realization of the 
high-spin particle, 
since the only off-shell particle in the matrix element 
$\langle v, p+q | T_{\mu\nu} | u, p \rangle$ is the graviton itself.

Our theorem {\em does} rely on one property of the graviton: its masslessness. 
If the graviton were massive, or if its propagator were modified in the 
infrared 
--as in~\cite{dgp} for instance-- then our theorem would not obtain. 
That alone is not sufficient to make gravitational interactions of 
high-spin massless fields 
consistent. Indeed, if the graviton was massive, one could integrate 
it out to obtain an 
effective theory valid for momenta lower than the graviton mass. 
The integration would unavoidably result in four-particle 
interactions involving the high-spin states. 
No example of consistent interactions of this type exists for
spin $s>2$. Indeed, theorems proving the opposite in fairly general 
cases have been 
already given in the literature~\cite{bbvd,bb,bbcl}.

Of course, massive particles of spin larger than two do exist and their 
gravitational interactions do obey the 
principle of equivalence. In the case of massive particles, 
spurious polarizations ($v_s$) become 
indistinguishable from physical longitudinal polarizations ($v_l$) 
at energies $E\gg m$\footnote{In renormalizable gauge theories this 
property is known as the {\em Goldstone Equivalence Theorem}~\cite{clt,cg}.}
\beq
v_l={1\over m} v_s + {\cal O}(m/E),
\eeq{60}
with $v_s$ given by eq.~(\ref{34}) for Fermions or eq.~(\ref{44}) for Bosons. 
Instead of signaling an inconsistency of the theory, now the non-decoupling of $v_s$ 
signals the onset of a strong coupling regime, since the matrix elements depend on 
inverse powers of the mass. The same property
ensures that the massless limit is singular, as announced in the Introduction. 
One could try to cure this
pathology by modifying the matrix elements by terms that explicitly depend on 
inverse powers of 
$m$~\cite{p,cpd}. Such terms do cancel mass singularities in 
$\langle v, p+q | T_{\mu\nu} | u_l, p \rangle$~\cite{p,cpd} but they also 
introduce 
additional singularities in previously regular matrix elements; namely 
$T_{\mu\nu}$ matrix elements between transverse states, i.e. 
states of highest helicity $\pm s$.

Massless particles in Anti de Sitter space-time are to all purposes 
indistinguishable from very light massive particles. 
The physical reason is that the curvature radius of AdS, $R_{AdS}$, 
acts as an IR cutoff effectively decoupling
particles with larger Compton wavelength. 
Technically, this can be seen in the absence of mass 
discontinuities in the $m\rightarrow 0$ limit of the massive 
propagator~\cite{p,gvn,dw}. 
In accordance with
this expectation and with the existence of interacting massive particles, 
theories of massless interacting
high-spin particles have been proposed~\cite{v}. 
Also in accordance with our expectations 
is the fact that
these theories become strongly interacting at $E\sim 1/R_{AdS}$, i.e. 
at the lowest energy for which one can localize a particle within the 
AdS Hubble radius. 
What this means for the ultimate viability of
such theories is yet to be properly understood.

\subsubsection*{A Limit on the Abelian Gauge Coupling of High-Spin Particles}
Our proof is easily adapted to constrain the coupling of charged massless
particles to $U(1)$ gauge fields. We derive here the constraints for 
Fermionic particles only, in order to 
spare the reader further tedium, and because this example already teaches 
us a most important lesson.
The most general helicity-conserving matrix element of a 
$U(1)$ current between on-shell spin $s$ states is
\beq
\langle v, p+q | J_\mu | u, p \rangle = \sum_{k=0}^n A^k \bar{v}^k \gamma_\mu u^k, 
\qquad s=n+1/2.
\eeq{61}
This matrix element is automatically conserved because both $u$ and $v$ obey the massless 
Dirac equation.
When the $U(1)$ gauge vector is massless, spurious polarizations decoupling 
requires
\beq
\langle v, p+q | J_\mu | u_s, p \rangle = q^2 \Delta_\mu(q),
\eeq{62}
with $\Delta_\mu(q)$ analytic at $q^2=0$. Substitution of the 
spurious polarization~(\ref{34}) 
into eq.~(\ref{61}) and eq.~(\ref{62}) gives the condition
\beq
A^{k+1} -(n-k){q^2\over 2} A^k = {\cal O}(q^2), \qquad k=0,...,n-1.
\eeq{63}
The charge of the high-spin state is defined~\cite{ww} by 
\beq
e=\lim_{q\rightarrow 0}A^n(q). 
\eeq{64}
Since all
coefficients $A^k$ must be less singular than $q^2$, the decoupling 
condition eq.~(\ref{63}) implies $e=0$ 
for $n\geq 1$ i.e. spin $s \geq 3/2$. This result is in accordance with 
supergravity, 
which indeed allows massless gravitini to have 
dipole and higher-multipole interactions in flat space, but not  nonzero $U(1)$
charges. Charged spin-3/2 fields require either a 
mass\footnote{Kaluza Klein gravitini have a charge
proportional to their mass: $e\propto m/M_{Pl}$.} or a cosmological 
constant~\cite{z}. 
The result obtained here is also weaker 
than our main result on gravitational coupling. 
Indeed, positivity of energy forbids the existence of a
particle with no energy but gravitational multipole coupling. Neutral massless 
particles with dipole or multipole
electromagnetic coupling are instead a rather mundane possibility. 
There is one final aspect of charged 
particle dynamics that is not captured by our analysis. Standard 
renormalization group analysis says that Abelian interactions are 
free in the IR, so 
the IR charge of {\em any} massless particle is always zero.
In a certain sense, our theorem rules out only a part of those 
theories that are 
already ruled out by the RG properties of unbroken Abelian gauge theories. 

\subsubsection*{}
We would like to conclude with a speculation. It seems that ``normal'' 
massless particles can exist only for spin not larger 
than two. On the other hand, it could be possible that high-spin fields 
do not obey some of the most basic properties 
of ``normal'' particles. Could it be that they 
do not obey the principle of equivalence, yet they still interact 
with gravity through gravitational multipoles, 
as neutral particles can do when coupled to $U(1)$ fields? At the level of
tri-linear interactions the answer is in 
the affirmative~\cite{bl}. 
Yet, the cubic vertex of~\cite{bl}, or any other vertex that may have been 
proposed in the literature, cannot be extended beyond cubic order:
Weinberg's theorem~\cite{w64} forbids it\footnote{We thank A. Nicolis for 
pointing this out to us.}. This is seen by applying Weinberg's factorization 
argument, reviewed in Section 2, to a vertex with two 
spin $s$ particles and two gravitons. In the limit that one of the two 
gravitons becomes soft, eq.~(\ref{5}) implies that {\em all} other particles
in the vertex must have the same gravitational charge, say $g_g=g_s=g'_s=1$. 
This is true when the soft graviton ends on the external hard graviton. 
In this case the identity
$g_g=1$ means simply that the graviton self-interacts 
in accordance with the principle of equivalence. On the other hand, when 
the soft graviton ends on either of the two spin $s$ lines, 
our general argument (and, of course, 
the explicit vertex in ref.~\cite{bl}) gives $g_s=g'_s=0$. 

So, if high-spin massless fields do interact at all with ``normal'' matter, 
they cannot couple to any of its local degrees of freedom. 
They would have to couple to 
unusual, global  degrees of freedom. This is not impossible since similar 
objects have already appeared in field theory. 
For instance singleton fields
in AdS, which carry no bulk degree of freedom~\cite{affs}; 
the graviton of 3-d gravity~\cite{w88} and BF fields in various dimensions 
(\cite{bbrt} and references therein), which also propagate no local degrees of 
freedom, etc. Maybe high-spin massless fields could constitute a 
new type of highly unusual, ``quasi-topological'' matter. Some positive hints
that this may be true come from the study of the massless limit of Witten's
open string field theory~\cite{kob}

\subsection*{Acknowledgments}
Work supported in part by NSF grants PHY-0245068 and PHY-0758032. Part of this work was also
supported by a Marie Curie Excellence Chair, contract MEXC-CT-2003-509748 (SAG@SNS).

\end{document}